\documentclass[aps,twocolumn,showpacs,amsmath,amssymb,groupedaddress]{revtex4}

\usepackage{mathrsfs}
\usepackage{graphicx}
\usepackage{CJK}

\def\hatd#1{\hat{#1}^\dagger}
\def\bra#1{\left\langle{#1}\right|}
\def\ket#1{\left|{#1}\right\rangle}
\def\braket#1#2{\left\langle{{#1}}\mathrel{\left|{\vphantom{{#1}{#2}}}\right.\kern-\nulldelimiterspace}{{#2}}\right\rangle}
\def\slfrac#1#2{\left.#1\middle/#2\right.}

\begin{document}


\title{Statistics-dependent quantum co-walking of two particles in one-dimensional lattices with nearest-neighbor interactions}

\author{Xizhou Qin$^{1}$}
\author{Yongguan Ke$^{1}$}
\author{Xiwen Guan$^{2,3}$}
\author{Zhibing Li$^{1}$}
\author{Natan Andrei$^{4}$}
\author{Chaohong Lee$^{1,}$}
\altaffiliation{Corresponding author. Email: chleecn@gmail.com}

\affiliation{$^{1}$State Key Laboratory of Optoelectronic Materials and Technologies, School of Physics and Engineering, Sun Yat-Sen University, Guangzhou 510275, China}
\affiliation{$^{2}$State Key Laboratory of Magnetic Resonance and Atomic and Molecular Physics, Wuhan Institute of Physics and Mathematics, Chinese Academy of Sciences, Wuhan 430071, China}
\affiliation{$^{3}$Department of Theoretical Physics, Research School of Physics and Engineering, Australian National University, Canberra ACT 0200, Australia}
\affiliation{$^{4}$Department of Physics, Rutgers University, Piscataway, New Jersey 08854, USA}

\date{\today}

\begin{abstract}
We investigate continuous-time quantum walks of two indistinguishable particles [bosons, fermions or hard-core bosons (HCBs)] in one-dimensional lattices with nearest-neighbor interactions.
The results for two HCBs are well consistent with the recent experimental observation of two-magnon dynamics [Nature \textbf{502}, 76 (2013)].
The two interacting particles can undergo independent- and/or co-walking depending on both quantum statistics and interaction strength.
Two strongly interacting particles may form a bound state and then co-walk like a single composite particle with statistics-dependent walk speed.
Analytical solutions for the scattering and bound states, which appear in the two-particle quantum walks, are obtained by solving the eigenvalue problem in the two-particle Hilbert space.
In the context of degenerate perturbation theory, an effective single-particle model for the quantum co-walking is analytically derived and the walk seep of bosons is found to be exactly three times of the ones of fermions/HCBs.
Our result paves the way for experimentally exploring quantum statistics via two-particle quantum walks.
\end{abstract}


\maketitle

\section{\label{sec1}Introduction}

Quantum walk (QW)~\cite{Aharonov1993,Kempe2003}, the quantum counterpart of classical random walk (CRW), is not only a fundamental phenomenon in quantum transport, but also a practical tool for developing quantum algorithms and implementing quantum computations.
In contrast to CRWs, which gradually approach to an equilibrium distribution, QWs spread ballistically if there is no decoherence.
The non-classical features of QWs offer versatile applications in quantum simulation~\cite{Schreiber2012}, quantum computation~\cite{Childs2009,Childs2013}, detection of topological states~\cite{Kitagawa2010+Kitagawa2010a,Kitagawa2012, Kraus2012+Verbin2013} and bound states~\cite{Kitagawa2012,Ahlbrecht2012,Fukuhara2013a}, and so on.

Up to now, single-particle QWs have been implemented with several experimental systems.
In those experiments, the roles of quantum walkers are taken by single particles such as neutral atoms~\cite{Karski2009}, atomic ions~\cite{Schmitz2009+Zahringer2010}, photons~\cite{Schreiber2010+Broome2010+Schreiber2011}, atomic spin impurities~\cite{Fukuhara2013}, and nuclear-magnetic-resonance systems~\cite{Du2003}.
Attribute to their superpositions and interference features, single-particle QWs yield an exponential speedup over CRWs~\cite{Childs2003}.
However, it has been demonstrated that such an exponential speedup can be also achieved by classical waves~\cite{Knight2003+Perets2008}.

In contrast, multi-particle QWs may have exotic non-classical correlations, which may bring new benefits to practical quantum technologies.
It has found that two-particle discrete QWs sensitively depend on the entanglement or correlations~\cite{Omar2006, Pathak2007}.
Naturally, the quantum statistical nature of two bosonic/fermionic walkers result in the emergence of bunching and anti-bunching in two-particle QWs, respectively~\cite{Omar2006}.
Moreover, multi-particle QWs can be used to implement universal quantum computations~\cite{Childs2013}.
By using linear~\cite{Hillery2010,Sansoni2012, Peruzzo2010+Lahini2010+Meinecke2013} and nonlinear photonic waveguide arrays~\cite{Solntsev2012,Lahini2012}, two-particle QWs have been implemented in several laboratories.
Exotic quantum correlations have been observed even in the absence of inter-particle interactions~\cite{Bromberg2009,Benedetti2012, Peruzzo2010+Lahini2010+Meinecke2013}.
Recently, the coexistence of free and bound states~\cite{Ganahl2012+Liu2013} has been observed via two-particle QWs of atomic spin-impurities in one-dimensional (1D) optical lattices~\cite{Fukuhara2013a}.

Although there are some studies on two-particle QWs involving quantum statistics and inter-particle interactions, most of them only consider either how quantum statistics affects the QWs of two non-interacting particles~\cite{Benedetti2012,Sansoni2012} or how inter-particle interaction affects the QWs of two interacting particles with a specific quantum statistics~\cite{Lahini2012,Fukuhara2013a}.
Up to now, there is still lacking a comprehensive study on how two-particle QWs depend on both quantum statistics and inter-particle interactions.
It is particularly interesting that how the quantum co-walking of two interacting particles \emph{quantitatively} depends on the quantum statistics of two walkers.
Here, the co-walking means that the two walkers are fully synchronized and walk as a single composite unity.

In this article, we investigate two-particle continuous-time QWs in 1D lattices with nearest-neighbor interactions.
We concentrate on analyzing quantum statistic affects in the QWs of two interacting particles.
We show the bunching/anti-bunching dynamics induced by the Bose/Fermi natures of quantum walkers, and systematically investigate the statistics-dependent quantum co-walking.
In addition to the numerical results, we derive an analytical model for the statistics-dependent quantum co-walking by employing degenerate perturbation theory.
We present both analytical and numerical results which are well consistent with each other.
Our analytical results give a \emph{quantitative} understanding of the quantum statistic effects in the quantum co-walking of two interacting indistinguishable particles.
In particular, our prediction on two hard-core bosonic walkers agrees with the experimental observation of the two-magnon dynamics~\cite{Fukuhara2013a}.
In the scenario of quantum-optical analogue~\cite{Longhi2009,Szameit2010}, our two-particle QWs can be experimentally verified by the light propagations in two-dimensional (2D) waveguide arrays~\cite{Szameit2009+Corrielli2013,Szameit2010}.

The structure of our article is as following.
In this section, we introduce background and motivation.
In Sec.~\ref{sec2}, we describe our models and discuss some key properties of them.
In Sec.~\ref{sec3}, we solve the eigenvalue problem in the two-particle Hilbert space and derive several analytical solutions for the two-particle eigenstates and their eigen-energies.
In Sec.~\ref{sec4}, we analyze the two-particle QWs under three different types of statistics: bosonic, fermionic, and hard-core bosonic ones.
In both position and momentum spaces, two-body correlations of bosonic and fermionic walkers show subtle bunching and anti-bunching signatures, respectively.
However, hard-core bosonic walkers show anti-bunching signature in the position space and bunching signature in the momentum space.
In Sec.~\ref{sec5}, we analytically derive the effective single-particle model for the co-walking of two quantum walkers under strong inter-particle interactions.
We discuss the implementation of our model and summary the results in Sec.~\ref{sec6}.

\section{Model}\label{sec2}

We consider QWs of two indistinguishable particles in 1D lattice system described by the following Hamiltonian with periodic boundary conditions (PBCs),
\begin{equation}
  \hat H=-J\sum_{l=-L}^{L}{\left(\hat a_l^\dagger\hat a_{l+1}+\mathrm{h.c.}\right)}+V\sum_{l=-L}^{L}{\hat n_l\hat n_{l+1}}.\label{Eq_Hamiltonian}
\end{equation}
Here, the total number of lattice sites is $L_t=2L+1$, $\hat a_l^\dagger$ ($\hat a_l$) creates (annihilates) a particle on the $l$-th lattice ($l=-L,\cdots,0,\cdots,L$), $\hat n_l = \hat a_l^\dagger \hat a_{l}$ is the particle number, $J$ is the nearest-neighbor hopping, and $V$ stands for the nearest-neighbor interaction.
Below we only discuss the Hamiltonian of attractive interaction $V<0$.

The two-particle (i.e. $\hat{N}=\sum_{l=-L}^{L}{\hat n_l}=2$) propagation in our systems represents a class of continuous-time two-particle QWs.
The continuous-time QWs can be generalized from the continuous-time CRWs~\cite{Kempe2003,Childs2003}.
A CRW on a graph is described by a matrix $\mathbf{M}$, which transforms the probability distribution (PD) $\mathbf{p}=\{p_l\}$ over a vertex set $\mathbf{v}=\{l\}$ (here $p_l$ is the probability of finding the walker at the $l$-th vertex).
The elements $\mathbf{M}_{ll'}$ give the jumping rate from the $l$-th vertex to the $l'$-th vortex.
The PD evolution of such a walk follows $\frac{\mathrm d}{\mathrm dt}\mathbf{p}(t)=-\mathbf{M}\mathbf{p}(t)$ and the solution is given as $\mathbf{p}(t)=e^{-\mathbf{M}t}\mathbf{p}(0)$ with the initial condition $\mathbf{p}(0)$.
In the quantum case, the matrix $\mathbf{M}$ is replaced by the so-called adjacency matrix $\mathbf{H}$~\cite{Farhi1998,Childs2009}, which generates an unitary evolution $e^{-i\mathbf{H}t}$ instead of $e^{-\mathbf{M}t}$.
Starting from an initial state $\ket{\psi_{\mathrm{ini}}}$, the quantum state evolves according to $\ket{\psi (t)}=e^{-i\mathbf{H}t}\ket{\psi_{\mathrm{ini}}}$, and the PD over the vortex set is given by the quantum projection $p_l=\left|\langle l |\psi(t)\rangle\right|^2$ with $\ket{l}$ denoting the quantum state of the walker localized at the $l$-th vortex.
In our system, the graph for the two walkers is the 1D lattice, and the Hamiltonian matrix $\hat{H}$ plays the role of the adjacency matrix.

We consider three typical types of commutation relations (CRs): bosonic, fermionic and hard-core bosonic ones.
The bosonic CRs read as $[\hat a_l,\hat a_k]=[\hat a_l^\dagger,\hat a_k^\dagger]=0$ and $[\hat a_l,\hat a_k^\dagger]=\delta_{l k}$.
The fermionic CRs obey $\{\hat a_l,\hat a_k\}=\{\hat a_l^\dagger,\hat a_k^\dagger\}=0$ and $\{\hat a_l,\hat a_k^\dagger\}=\delta_{l k}$.
The hard-core bosonic CRs are described by $[\hat a_l,\hat a_k]=[\hat a_l^\dagger,\hat a_k^\dagger]=[\hat a_l,\hat a_k^\dagger]=0$ for $l\neq{k}$, while $\{\hat a_l,\hat a_l\}=\{\hat a_l^\dagger,\hat a_l^\dagger\}=0$ and $\{\hat a_l,\hat a_l^\dagger\}=1$.

The Hamiltonian~(\ref{Eq_Hamiltonian}) associates with the quasi-particle representation for an XXZ Heisenberg chain~\cite{Matsubara1956,Jordan1928+Dziarmaga2005}.
By using the mapping: $\ket{\downarrow}\leftrightarrow\ket{0}$, $\ket{\uparrow}\leftrightarrow\ket{1}$, $\hat S_l^+\leftrightarrow\hatd a_l$, $\hat S_l^-\leftrightarrow\hat a_l$ and $\hat S_l^z\leftrightarrow\hat n_l-{1 \over 2}$, the hard-core bosonic system is equivalent to the XXZ Heisenberg chain~\cite{Matsubara1956},
\begin{eqnarray}
  \hat H_\mathrm{XXZ}&=&-J_\mathrm{ex}\sum_l\left(\hat S_l^x\hat S_{l+1}^x+\hat S_l^y\hat S_{l+1}^y+\Delta\hat S_l^z\hat S_{l+1}^z\right) \nonumber \\
  && +h_z\sum_l\hat S_l^z,\label{Eq_XXZModel}
\end{eqnarray}
with $J_\mathrm{ex}=2J$, $\Delta=-{V\over{2J}}$, $h_z=V$ and $\hat S_l^{\pm}=\hat S_l^x\pm i\hat S_l^y$.
It has been demonstrated that such an XXZ Heisenberg chain can be realized by ultracold two-level atoms in optical lattices~\cite{Duan2003+Kuklov2003+Garcia-Ripoll2003+Altman2003+Lee2004, Fukuhara2013, Fukuhara2013a}.

\section{Two-particle eigenstates}\label{sec3}

In this section, we solve the eigenvalue problem in the two-particle Hilbert space and give the eigenstates which appear in the two-particle QWs.
Since $[\hat{N}, \hat{H}]=0$, the total particle number $\hat{N}$ is conserved and all initial two-particle states keep evolving in the two-particle Hilbert space.
For two bosons, the Hilbert space is spanned by basis,
$$
\mathcal{B}^{(2)}_{\mathrm{B}}=\left\{\ket{l_1 l_2} = (1+\delta_{l_1 l_2})^{-\frac{1}{2}} \hatd a_{l_1}\hatd a_{l_2}\ket{\mathbf{0}}\right\},
$$
with $-L\leq l_1 \leq l_2\leq L$.
For two fermions or two hard-core bosons (HCBs), the Hilbert spaces are spanned by basis,
$$
\mathcal{B}^{(2)}_{\mathrm{FH}}=\left\{\ket{l_1 l_2}=\hatd a_{l_1}\hatd a_{l_2}\ket{\mathbf{0}}\right\},
$$
with $-L \leq l_1<l_2 \leq L$.
Given $\mathcal{B}^{(2)}_{\mathrm{B}}$ and $\mathcal{B}^{(2)}_{\mathrm{FH}}$, it is easy to find the Hamiltonian matrix $H^{(2)}$ in the two-particle sector.

Introducing $C_{l_1 l_2}=\bra{\mathbf{0}}\hat a_{l_2}\hat a_{l_1}\ket{\Psi}$, the eigenstates can be expanded as $\ket{\Psi}=\sum_{l_1 \leq l_2}\psi_{l_1 l_2}\ket{l_1 l_2}$ with $\psi_{l_1 l_2}=C_{l_1 l_2}(1+\delta_{l_1 l_2})^{-\frac{1}{2}}$.
Independent upon the quantum statistics of particles, the eigenequation $\hat H\ket{\Psi}=E\ket{\Psi}$ can be written in a unified form of
\begin{eqnarray}\label{Eq_Eigene_Quation}
  EC_{l_1 l_2}&=&-J\left(C_{l_1,l_2+1}+C_{l_1,l_2-1}+C_{l_1+1,l_2}+C_{l_1-1,l_2}\right) \nonumber \\
  &&+V\delta_{l_1,l_2\pm 1}C_{l_1 l_2},
\end{eqnarray}
with $\delta_{l_1,l_2\pm 1}=1$ if $l_1=l_2\pm 1$ and $\delta_{l_1,l_2\pm 1}=0$ if $l_1 \ne l_2\pm 1$.
Here, the PBC requires $C_{l_1+L_t,l_2}=C_{l_1,l_2+L_t}=C_{l_1 l_2}$.
The CRs require that $C_{l_1 l_2}=C_{l_2 l_1}$ for bosons; $C_{l_1 l_1}=0$ and $C_{l_1 l_2}=-C_{l_2 l_1}$ for fermions; and $C_{l_1 l_1}=0$ and $C_{l_1 l_2}=C_{l_2 l_1}$ for HCBs.

The motion of the two-particle system can be separated by the motion of the center-of-mass $R=\frac{1}{2}(l_1+l_2)$ and the one of the relative position $r=l_1-l_2$.
By employing the ansatz $C_{l_1 l_2}=e^{i K R}\phi(r)$, the eigenequation reads as
\begin{equation}
  E\phi(r)=J_K\left(\phi(r-1)+\phi(r+1)\right)+V\delta_{r,\pm 1}\phi(r) \label{Eq_Relative_Motion}
\end{equation}
with $J_K=-2J\cos\left(\frac{K}{2}\right)$, $\delta_{r,\pm 1}=1$ if $r=\pm1$ and $\delta_{r,\pm 1}=0$ if $r \ne \pm1$.
Therefore, the PBC requires $e^{i K L_t}=1$ and $\phi(r+L_t)=e^{i K L_t/2}\phi(r)$ with the quantized total quasi-momentum $K=2\pi\alpha/L_t$ with $\alpha=-L,-L+1,\cdots,L$.
Correspondingly, the CRs require that $\phi(r)=\phi(-r)$ for bosons; $\phi(0)=0$ and $\phi(r)=-\phi(-r)$ for fermions; $\phi(0)=0$ and $\phi(r)=\phi(-r)$ for HCBs.

The PBC and CRs indicate that $\{\phi(r)|r=0,\cdots,L\}$ for bosons and $\{\phi(r)|r=1,\cdots,L\}$ for fermions/HCBs are independent variables.
Thus the two-particle Hamiltonian matrix block for bosons with total quasi-momentum $K$ can be written as
\begin{eqnarray}
  \hat H^{(2)}_\mathrm{B}(K)=\left(\begin{array}{cccccc}
     0   &  2J_K &   \   &   \   &   \   &   \   \\
    J_K  &   V   &  J_K  &   \   &   \   &   \   \\
     \   &  J_K  &   0   &  J_K  &   \   &   \   \\
     \   &   \   &\ddots &\ddots &\ddots &   \   \\
     \   &   \   &   \   &  J_K  &   0   &  J_K  \\
     \   &   \   &   \   &   \   &  J_K  &  J_K^\mathrm{B} \end{array}\right),&&\label{Eq_Hamiltonian_Boson}
\end{eqnarray}
the one for fermions reads as
\begin{eqnarray}
  \hat H^{(2)}_\mathrm{F}(K)=\left(\begin{array}{ccccc}
     V   &  J_K  &   \   &   \   &   \   \\
    J_K  &   0   &  J_K  &   \   &   \   \\
     \   &\ddots &\ddots &\ddots &   \   \\
     \   &   \   &  J_K  &   0   &  J_K  \\
     \   &   \   &   \   &  J_K  &  J_K^\mathrm{F} \end{array}\right),&&\label{Eq_Hamiltonian_Fermion}
\end{eqnarray}
and the one for HCBs is in form of
\begin{eqnarray}
\hat H^{(2)}_\mathrm{H}(K)=\left(\begin{array}{ccccc}
     V   &  J_K  &   \   &   \   &   \   \\
    J_K  &   0   &  J_K  &   \   &   \   \\
     \   &\ddots &\ddots &\ddots &   \   \\
     \   &   \   &  J_K  &   0   &  J_K  \\
     \   &   \   &   \   &  J_K  &  J_K^\mathrm{H} \end{array}\right).&&\label{Eq_Hamiltonian_HCB}
\end{eqnarray}
Here, we define $J_K^\mathrm{H}=J_K^\mathrm{B}=-J_K^\mathrm{F}=e^{i K L_t/2} J_K$.

The Hamiltonian matrices~(\ref{Eq_Hamiltonian_Boson}, \ref{Eq_Hamiltonian_Fermion}, \ref{Eq_Hamiltonian_HCB}) can be diagonalized numerically and analytically~\cite{Scott1994+Valiente2008+Valiente2009+Nguenang2009}.
When $V\ne 0$, for all three cases (bosons, fermions and HCBs), there are two types of eigenstates: bound states (BSs) and scattering states (SSs).
For SSs, the amplitude of the wave function, $\phi(r)$, oscillates as the relative position $r$, while for BSs, it decays exponentially.
The general eigenstate for Eq.~(\ref{Eq_Relative_Motion}) can be expressed as
\begin{equation}
  \phi(r)=A_+e^{ikr}+A_-e^{-ikr}
\end{equation}
with the two constants ($A_+$, $A_-$) and the quasi-momentum $k$.
For SSs, the quasi-momentum $k$ is real. However, for BSs, the quasi-momentum $k$ is purely imaginary.
The eigenenergy is given as
\begin{equation}
E^{(2)}_{K,k}=2 J_K\cos(k) =-4J\cos\left(\frac{K}{2}\right)\cos(k) \label{EE}
\end{equation}
with the quasi-momentum $k$ determined by the physical parameters and the statistical properties. Below, we will show how to determine the quasi-momentum $k$.

\emph{(A) Scattering States.} -
Due to the real value of $k$, the scattering states are invariant under the transformation: $k\rightarrow k\pm 2\pi$ and $k\rightarrow -k$. Thus we only need consider $0\leq k<\pi$.

For fermions, the PBC and CRs require that
$$
\left\{\begin{array}{rcc}
  (J_K-Ve^{i k})A_+ + (J_K-Ve^{-i k})A_- &=& 0, \\
  e^{i k L_t}A_+ + (-1)^\alpha A_- &=& 0.
\end{array}\right.
$$
By eliminating $A_+$ and $A_-$, one can obtain that the quasi-momentum $k$ obeys
\begin{equation}\label{Eq_k_Fermion}
  e^{i k L_t}=(-1)^\alpha{{J_K-Ve^{i k}}\over{J_K-Ve^{-i k}}}.
\end{equation}
To give all possible values of $k$, one has to solve Eq.~(\ref{Eq_k_Fermion}), which is actually an algebraic equation of $e^{i k}$.
Thus, the corresponding eigenstate reads as
\begin{equation}
  \phi(r)\propto e^{ikr}-e^{-ikr}e^{i\left({K\over 2}+k\right)L_t}, (1\leq r\leq N).
\end{equation}

For HCBs, the quasi-momentum $k$ satisfies
\begin{equation}\label{Eq_k_HCB}
  e^{i k L_t}=(-1)^{\alpha-1}{{J_K-Ve^{i k}}\over{J_K-Ve^{-i k}}},
\end{equation}
and the corresponding eigenstate is
\begin{equation}
  \phi(r)\propto e^{ikr}+e^{-ikr}e^{i\left({K\over 2}+k\right)L_t}, (1\leq r\leq N).
\end{equation}

For bosons, the ansatz should be modified as
\begin{equation}
  \phi(r)=\left\{\begin{array}{ll}
  \phi_0, & r=0, \\
  A_+e^{ikr}+A_-e^{-ikr}, & 1\leq r\leq N,\end{array}\right.
\end{equation}
with the quasi-momentum $k$ satisfying
\begin{equation}\label{Eq_k_Boson}
  e^{ikL_t} =(-1)^{\alpha}{{J_K(e^{ik}-e^{-ik}) +V(1+e^{2ik})}\over{J_K(e^{ik}-e^{-ik})-V(1+e^{-2ik})}}.
\end{equation}
Thus the corresponding eigenstate is given as
\begin{equation}
  \phi(r)\propto e^{ikr}+e^{-ikr}e^{i\left({K\over 2}+k\right)L_t}, (1\leq r\leq N),
\end{equation}
with $\phi(0)=\phi_0=\phi(1)/\cos(k)$.

\emph{(B) Bound States.} - The bound states correspond to purely imaginary $k=i\eta$ ($\eta>0$) satisfying the conditions~(\ref{Eq_k_Fermion}, \ref{Eq_k_HCB}, \ref{Eq_k_Boson}).
For a finite $L_t$, no compact formulae for $\eta$ are available.
But when $L_t$ is sufficiently large, the factor $e^{i k L_t}=e^{-\eta L_t}$ become small, as an approximation, one can assume $e^{-\eta L_t}\approx 0$, which is exact ($e^{-\eta L_t}=0$) when $L_t\rightarrow \infty$.
Then the conditions for $\eta$ read as
\begin{equation}\label{Eq_k}
  \left\{\begin{array}{ll}
    J_K=Ve^{-\eta}, & \mathrm{fermions/HCBs}, \\
    J_K(e^{-\eta}-e^{\eta})+V(1+e^{-2\eta})=0, & \mathrm{bosons}.
  \end{array}\right.
\end{equation}
Solving the above equation, one can obtain
\begin{equation}
e^{\eta}=V/J_K \label{eta_FermiHCB}
\end{equation}
for fermions/HCBs (as long as $|V|>|J_K|$), and
\begin{eqnarray}
  e^{\eta}&=&{1 \over 3}\left(\beta+{{3+\beta^2} \over \Delta_0}+\Delta_0\right),\label{eta_Bose} \\
  \Delta_0&=&\left(18\beta +\beta^3 +3\sqrt{3}\sqrt{\beta^4+11\beta^2 -1}\right)^{1/3}, \nonumber
\end{eqnarray}
for bosons (as long as $\beta^2(\beta^2+11)>1$), where $\beta=V/J_K$.

According to Eq.~(\ref{EE}), given $k=i\eta$ for a BS, its eigenenergy reads as
\begin{equation}
E^{(2)}_{K,\eta}=2J_K\cosh(\eta)\label{EE_BS}
\end{equation}
with $\eta$ determined by Eq.~(\ref{Eq_k}).
Thus for the BS of fermions or HCBs, according to Eq.~(\ref{eta_FermiHCB}), its eigenenergy reads as
\begin{equation}
  E^{(2)}_\mathrm{FH}(K)=V+{4J^2 \over V}\cos^2\left({K \over 2}\right), \label{Eq_FH_E_Exact}
\end{equation}
when $\left|V/(2J)\right|>1$ and $L_t\rightarrow\infty$.
Obviously, Eq.~(\ref{Eq_FH_E_Exact}) fully agree with the ones obtained from the Bethe ansatz~\cite{Takahashi1999}.
For the case of strongly interacting bosons, that is $\left|\beta\right| \rightarrow \infty$, we have $e^{\eta}>\beta=V/J_K$ and $e^{\eta}/\beta\rightarrow 1$ for positive $\beta$.
Thus according to Eqs.~(\ref{eta_FermiHCB},\ref{eta_Bose},\ref{EE_BS}), for the case of attractive interaction, the BS eigenenergy of bosons is lower than the one of fermions/HCBs, and their difference vanishes when $\left|V/(2J)\right|\rightarrow\infty$.

\begin{figure}[!htp]
  \includegraphics[width=1.0\columnwidth]{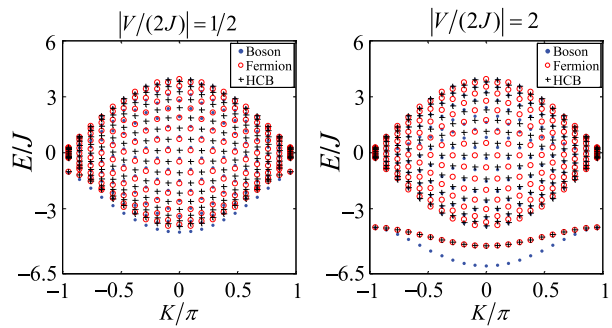}
  \caption{\label{Fig_Spectrum}(Color online) Two-particle spectrum for a 21-site system~(\ref{Eq_Hamiltonian}) with different interaction strength $\left|V/2J\right|$. Left: weak interaction, $\left|V/2J\right|=1/2$. Right: strong interaction, $\left|V/2J\right|=2$. Each point represents an eigenenergy $E$ for a given quasi-momentum $K$.}
\end{figure}

In Fig.~\ref{Fig_Spectrum}, we show the energy spectrum for the two-particle system. For weak interaction, $\left|V/(2J)\right|<1$, there is only one band, in which SSs and BSs are mixed. For strong interaction, $\left|V/(2J)\right|>1$, there are two mini-bands, in which the upper band corresponds to SSs and the lower band corresponds to BSs.

\section{Two-particle quantum walks}\label{sec4}

In this section, we focus on the time-evolution dynamics of two-particle states, i.e., the two-particle QWs.
In particular, by analyzing two-particle correlations in both position and momentum spaces, we explore how interaction and statistics affect the two-particle QWs.

In units of $\hbar=1$, the two-particle QWs obeys the time-dependent Schr\"odinger equation
\begin{equation}
  i\frac{\mathrm{d}}{\mathrm{d}t}\left|\psi(t)\right\rangle =H^{(2)} \left|\psi(t)\right\rangle,\label{Eq_TE}
\end{equation}
with $\left|\psi(t)\right\rangle=\sum_{l_1 \le l_2} \psi_{l_1 l_2} (t) \ket{l_1 l_2}$ for bosons and $\left|\psi(t)\right\rangle=\sum_{l_1 < l_2} \psi_{l_1 l_2} (t) \ket{l_1 l_2}$ for fermions and HCBs.
Here we consider the two-particle QWs from an initial state of two particles siting in neighbouring lattice sites, $\left|\psi_\mathrm{ini}\right\rangle=\hat a_0^\dagger\hat a_1^\dagger\left|\mathbf{0}\right\rangle$. Here, $\ket{\mathbf{0}}$ denotes the vacuum state.

\begin{figure*}[!htp]
\includegraphics[width=2.0\columnwidth]{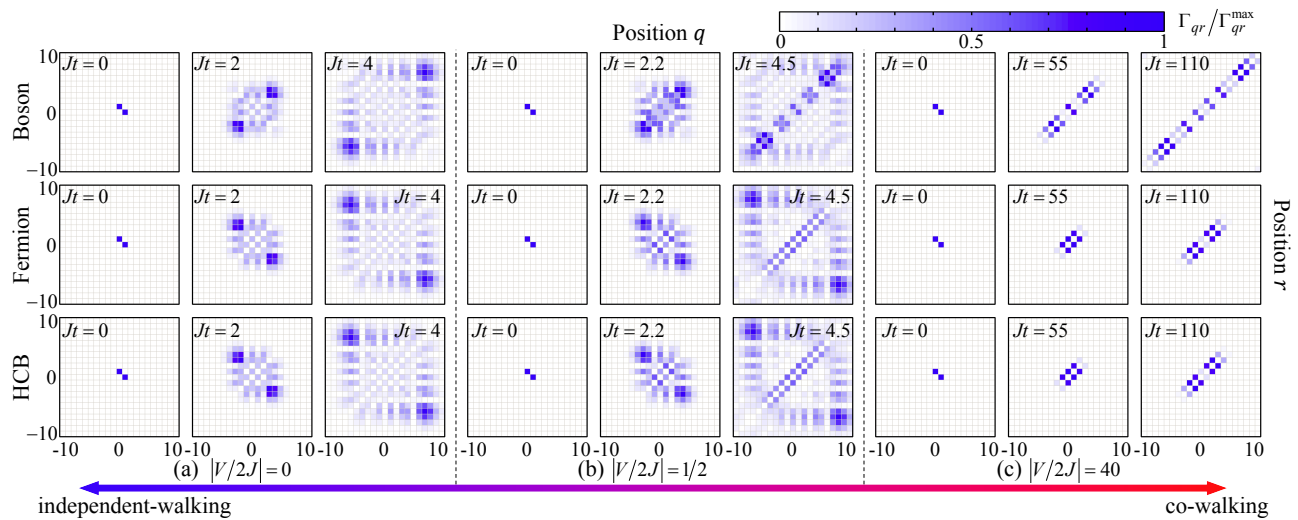}
\caption{\label{Fig_TPCF_Position}(Color online) Two-particle correlations of quantum walkers in position space. The first, second and third rows correspond to Bose, Fermi and HCB statistics, respectively. The interaction-hopping ratios $\left|V/(2J)\right|$ are (a) $0$, (b) $0.5$, and (c) $40$. Here, the total number of lattice sites is $L_t=21$, the evolution time is given by $Jt$, and we only show the instantaneous correlations before the particles collide with the boundaries $l=\pm 10$.}
\end{figure*}

\begin{figure*}[!htp]
\includegraphics[width=2.0\columnwidth]{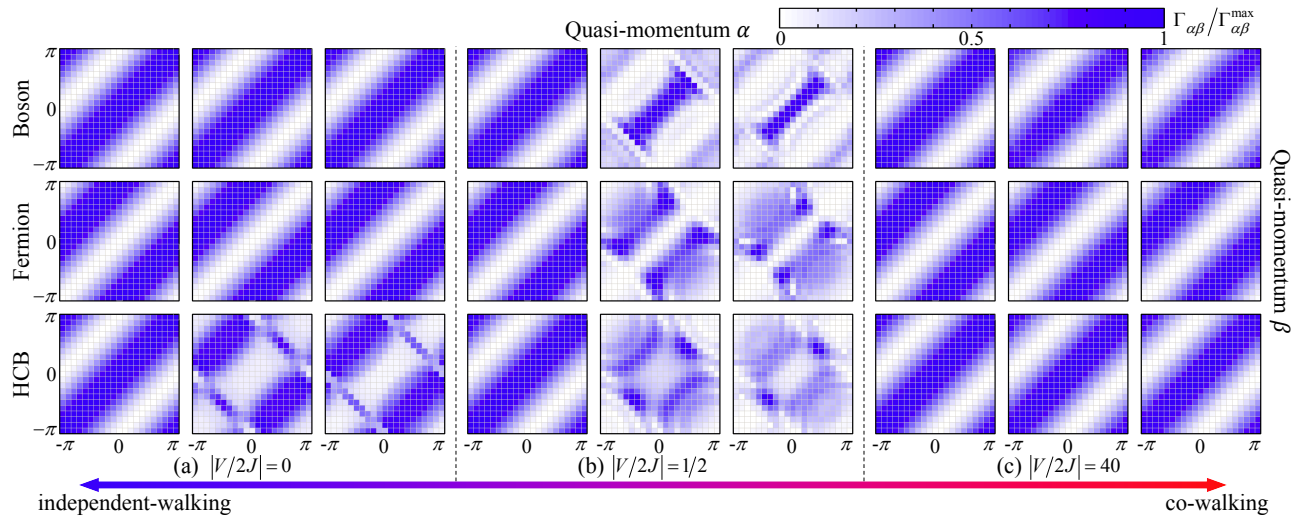}
\caption{\label{Fig_TPCF_Momentum}(Color online) Two-particle correlations of quantum walkers in momentum space with the same setting of Fig.~\ref{Fig_TPCF_Position}.}
\end{figure*}

To explore the correlation between two quantum walkers, we calculate the time-dependent two-particle correlation in position space,
\begin{equation}
  \Gamma_{q r}(t)=\bra{\psi(t)}\hatd a_q\hatd a_r\hat a_r\hat a_q\ket{\psi(t)},
\end{equation}
and the ones in momentum space,
\begin{equation}
  \Gamma_{\alpha \beta}(t)=\bra{\psi(t)}\hatd c_\alpha\hatd c_\beta\hat c_\beta\hat c_\alpha\ket{\psi(t)},
\end{equation}
with $\ket{\psi(t)}$ in Eq.~(\ref{Eq_TE}).
Here, $\hat c_\alpha^\dagger={1 \over \sqrt{L_t}} \sum_{l=-L}^{L}e^{-ip_\alpha l}\hat a_l^\dagger$ is the discrete Fourier transformation of $\hatd a_l$, in which the quasi-momentum $p_\alpha=2\pi\alpha/L_t$, the integer $\alpha=-L, -L+1,\cdots,L$.
The two-particle correlation in position and momentum spaces for different quantum statistics and interaction strength provide a clear insight into the two-particle QWs, see Figs.~\ref{Fig_TPCF_Position} and \ref{Fig_TPCF_Momentum} for a 21-site system.
For systems with same parameters but different lattice sizes $L_t$, our numerical results show that, before the two particles collide with the boundaries, the finite-size effect and the boundary effect are negligible.

It is possible to distinguish co-walking from independent walking through examining the evolution of the spatial correlation $\Gamma_{qr}(t)$.
The co-walking of the two particles is signatured as the significant correlations at two specific lines ($q=r \pm d$) in the $(q,r)$-plane, where $d$ is a fixed integer dependent on the form of the inter-particle interaction.
For our systems with the nearest-neighbor interactions, the spatial correlation peaks (i.e. the peaks of $\Gamma_{qr}(t)$) appear on the two minor-diagonal lines ($q=r \pm 1$). This is a typical signature of the two-particle co-walking.

In the position space, the correlations of two bosonic walkers (the first row of Fig.~\ref{Fig_TPCF_Position}) show bunching behavior, while the correlations of two fermionic walkers (the second row of Fig.~\ref{Fig_TPCF_Position}) and two hard-core bosonic walkers (the third row of Fig.~\ref{Fig_TPCF_Position}) show anti-bunching behavior.
We observe that the correlations of fermions and HCBs in the position space almost have no difference.
This is because that, a spin-$\frac{1}{2}$ Heisenberg XXZ model, which is equivalent to a hard-core Bose-Hubbard model~\cite{Matsubara1956}, can be exactly mapped onto a Hubbard-like model of spinless fermions via Jordan-Wigner transformation~\cite{Jordan1928+Dziarmaga2005}.
Although boundary conditions for the Hubbard-like model of spinless fermions depend on the total particle number~\cite{Jordan1928+Dziarmaga2005}, before the two walkers hit the boundaries, the boundary condition effect on the two fermionic walkers is as same as the one on the two hard-core bosonic walkers.
Therefore the correlations are almost the same for fermions and HCBs in position space.

On the other hand, the correlations of bosonic and hard-core bosonic walkers in momentum space show bunching behavior, see the first and third rows of Fig.~\ref{Fig_TPCF_Momentum}.
Nevertheless the correlations of fermionic walkers (the second row of Fig.~\ref{Fig_TPCF_Momentum}) show anti-bunching behavior.
This means that bunching and anti-bunching in momentum space can show the difference between fermions and HCBs.
Therefore, bunching and anti-bunching of the two quantum walkers in both position and momentum spaces completely reveal the difference among bosons, fermions and HCBs.

The spatial correlations $\Gamma_{qr}$ on the minor diagonal lines $(q=r \pm 1)$ are gradually enhanced when the interaction-hopping ratio increases, see Fig.~\ref{Fig_TPCF_Position}.
Since $\Gamma_{q,q\pm 1}$ presents a joint probability of finding one walker on the $q$-th site and the other walker on the $(q\pm 1)$-th site, the significant correlations on the minor diagonal lines is a robust signature of quantum co-walking.
The quantum co-walking is also an important signature of the existence of two-particle bound states, see~\cite{Fukuhara2013a, Ganahl2012+Liu2013} for the case of two magnons.
Detailed discussions on quantum co-walking will be presented in the next section.
Usually, two interacting quantum walkers simultaneously undergo independent- and co-walking when the interaction is not strong enough.

\section{Effective dynamics of two-particle quantum co-walking}\label{sec5}

In this section, we will analytically derive an effective single-particle model for the quantum co-walking of two interacting particles and discuss the statistics-dependent behavior of quantum co-walking.
To our best knowledge, for the first time, we present a quantitative description of the quantum statistics effect in two-particle QWs.

Under strong inter-particle interactions ($\left|V/J\right| \gg 1$), the two quantum walkers behave as a single composite particle and their QWs are dominated by quantum co-walking.
As $\left|V/J\right| \gg 1$, one thus can treat the hopping term
\begin{equation}
\hat{H}_1 = -J\sum_{l=-L}^{L}{\left(\hat{a}_{l}^{\dagger} \hat {a}_{l+1} +\mathrm{h.c.}\right)}
\end{equation}
as a perturbation to the interaction term,
\begin{equation}
\hat H_0=V\sum_{l=-L}^{L}{\hat n_l\hat n_{l+1}},
\end{equation}
in the considered Hamiltonian~(\ref{Eq_Hamiltonian}).
By employing the second-order perturbation theory for degenerate systems~\cite{Takahashi1977}, we analytically obtain an effective single-particle model for the co-walking of the two quantum walkers.

To implement the perturbation analysis, we should give the projection operator onto the subspace involved the quantum co-walking and the projection operator onto the orthogonal component of the involved subspace.
The unperturbed Hamiltonian $\hat H_0$ has only two eigenvalues: (i) $E_0=V (<0)$ for the $L_t$-fold degenerated ground-states $\{\ket{G_{l}}=\ket{l,l+1}:-L\leq l\leq L\}$, and (ii) $E_{l_1 l_2}=0$ for excited eigenstates $\{\ket{E_{l_1 l_2}}=\ket{l_1,l_2}:l_1\neq l_2\pm 1\ {\mathrm{and}}\ -L\leq l_1\le l_2\leq L \mathrm{\ for\ bosons\ while\ } -L\le l_1<l_2 \leq L \mathrm{\ for\ fermions\ and\ HCBs}\}$.
The quantum co-walking only involves the subspace spanned by $L_t$ independent ground-states $\{\ket{G_{l}}\}$.
Denoting $\mathcal{U}_0=\{\ket{G_{l}}\}$, the projection operator onto $\mathcal{U}_0$ is
$$\hat P_0=\sum_l\ket{G_l}\bra{G_l}.$$
Introducing $\mathcal{V}_0$ as the orthogonal complement of $\mathcal{U}_0$, the projection onto $\mathcal{V}_0$ reads as
$$\hat S=\sum_{E_{l_1 l_2}\ne E_0}{\frac{1}{E_0-E_{l_1 l_2}}\ket{E_{l_1 l_2}}\bra{E_{l_1 l_2}}}.$$

Therefore, the effective Hamiltonian up to $2^\mathrm{nd}$ order is given as
\begin{equation}
  \hat H^{(2)}_\mathrm{eff}=\hat h_0+\hat h_2=E_0\hat P_0+\hat P_0\hat H_1\hat S\hat H_1\hat P_0.
\end{equation}
Since $E_{l_1 l_2}=0$, we have
\begin{eqnarray}
  \hat h_2=\frac{J^2}{V}\sum_{ll'jj'l_1l_2}
  \Big[&&\ket{G_l}\bra{G_l}\hat{T}_j\ket{E_{l_1 l_2}} \nonumber \\
  &&\times\bra{E_{l_1 l_2}}\hat{T}_{j'}\ket{G_{l'}}\bra{G_{l'}}\Big],
\end{eqnarray}
where the hoping operators $\hat{T}_k=\hatd a_k\hat a_{k+1}+\hatd a_{k+1}\hat a_k$, the summation indices $\{l, l', j, j'\}$ take values from $\{-L,-L+1,\cdots,L\}$, and $\{l_1,l_2\}$ is summed over all states of $E_{l_1 l_2}=0$.

Introducing the following two notations,
\begin{eqnarray}
  &&T^{jl}_{l_1 l_2}=\bra{G_l}(\hatd a_j\hat a_{j+1}+\hatd a_{j+1}\hat a_j)\ket{E_{l_1 l_2}}, \\
  &&\ket{G'_{l_1 l_2}}=\sum_{jl}{T^{jl}_{l_1 l_2}\ket{G_l}},
\end{eqnarray}
we have $\bra{G'_{l_1 l_2}}=\sum_{j'l'}{\bra{G_{l'}}T^{j'l'*}_{l_1 l_2}}$, and
\begin{eqnarray}
  \hat h_2&=&{{J^2} \over V}\sum_{l l' j j' l_1 l_2}\ket{G_l}T^{jl}_{l_1 l_2}T^{j'l'*}_{l_1 l_2}\bra{G_{l'}} \nonumber \\
  &=&{{J^2} \over V}\sum_{l_1 l_2}\ket{G'_{l_1 l_2}}\bra{G'_{l_1 l_2}} \label{Eq_h2}.
\end{eqnarray}
By using the CRs and $\hat a_l\hatd a_k\ket{\mathbf{0}} =\delta_{lk}\ket{\mathbf{0}}$,
after some algebra, we obtain
\begin{eqnarray}\label{G_l1l2}
  \ket{G'_{l_1 l_2}}&=&
  \sqrt 2\delta_{l_1 l_2} \left(\ket{G_{l_1-1}}+\ket{G_{l_1}}\right) \nonumber \\
  &&+\delta_{l_1,l_2-2} \left(\ket{G_{l_1}}+\ket{G_{l_1+1}}\right) \nonumber \\
  &&+\epsilon'\delta_{l_1-2,l_2} \left(\ket{G_{l_2}}+\ket{G_{l_2+1}}\right).
\end{eqnarray}
Here, $\epsilon'=1$ for bosons and HCBs, while $\epsilon'=-1$ for fermions.
Inserting Eq.~(\ref{G_l1l2}) into Eq.~(\ref{Eq_h2}), we get
\begin{eqnarray}
  \hat h_2=\frac{J^2}{V}&&\sum\limits_{l_1 l_2}\Big[
  2\delta_{l_1 l_2}\left(\ket{G_{l_1-1}}+\ket{G_{l_1}}\right) \left(\bra{G_{l_1-1}}+\bra{G_{l_1}}\right) \nonumber \\
  &&+\delta_{l_1,l_2-2}\left(\ket{G_{l_1}}+\ket{G_{l_1+1}}\right) \left(\bra{G_{l_1}}+\bra{G_{l_1+1}}\right) \nonumber \\
  &&+\delta_{l_1-2,l_2}\left(\ket{G_{l_2}}+\ket{G_{l_2+1}}\right) \left(\bra{G_{l_2}}+\bra{G_{l_2+1}}\right)\Big].\nonumber \\
\end{eqnarray}

For the case of fermions or HCBs, $\delta_{l_1 l_2}=0$ and $\delta_{l_1,l_2-2}=1$ for $(l_1,l_2)=(l,l+2)$ with $l=(-L, -L+1, \cdots, L-2)$ and $\delta_{l_1-2,l_2}=1$ for $(l_1,l_2)=(-L,L-1)$ and $(-L+1,L)$, thus we have
\begin{equation}
  \hat{h}_2 = {{J^2} \over V}\sum_{q=-L}^L \left(\ket{G_q}+\ket{G_{q+1}}\right) \left(\bra{G_q}+\bra{G_{q+1}}\right). \label{Eq_Eff_FH}
\end{equation}

For the case of bosons, besides the terms included in the case of fermions or HCBs, $\delta_{l_1 l_2}=1$ for $(l_1,l_2)=(l, l)$ with $l=(-L,-L+1,\cdots,L-1,L)$ should be included, thus we have
\begin{equation}\label{Eq_Eff_B}
  \hat h_2=\frac{3J^2}{V}\sum_{q=-L}^L\left(\ket{G_q}+\ket{G_{q+1}}\right) \left(\bra{G_q}+\bra{G_{q+1}}\right).
\end{equation}

In our model of nearest-neighbor-interaction, for two walkers starting from two neighbor lattice sites, their co-walking can be described by superposition of multiple ground states $\ket{G_q}=\hatd a_q\hatd a_{q+1}\ket{\mathbf{0}}=\ket{n_q=1,n_{q+1}=1}$ with different $q$ (where $q=-L,-L+1,\cdots,L-1,L$).
During the process of co-walking, the two particles behave like a single composite particle.

In order to capture the single-particle nature of the co-walking, we introduce creation operators $\hatd b_q$ for the composite particle consisting of one particle on the $q\textrm{-th}$ lattice site and the other particle on the $\left(q+1\right)\textrm{-th}$ lattice site.
Explicitly, $\hatd b_q \Leftrightarrow \hatd a_q\hatd a_{q+1}$ and $\ket{n^c_q=1}=\hatd b_q\ket{\mathbf{0}} \Leftrightarrow \ket{n_q=1,n_{q+1}=1}=\hatd a_q\hatd a_{q+1}\ket{\mathbf{0}}$.
Then, from Eq.~(\ref{Eq_Eff_B}), the two bosonic walkers obey an effective single-particle Hamiltonian,
\begin{equation}
  \hat H^{\mathrm{B}}_\mathrm{eff} = J^\mathrm{B}_\mathrm{eff} \sum_q\left(\hatd b_q\hat b_{q+1} +\hatd b_{q+1}\hat b_q\right)+\mu^\mathrm{B}_\mathrm{eff} \sum_q\hatd b_q\hat b_q, \label{Eq_Effective_b}
\end{equation}
with the hopping strength $J^\mathrm{B}_\mathrm{eff}=\slfrac{3J^2}{V}$ and the chemical potential $\mu^\mathrm{B}_\mathrm{eff}=V+\slfrac{6J^2}{V}$.
The spectrum of Hamiltonian~(\ref{Eq_Effective_b}) can be obtained by substituting the ansatz $\ket{\psi}=\sum_me^{iKm}\hatd b_m\ket{\mathbf{0}}$ into the eigenvalue problem $\hat H^{B}_\mathrm{eff}\ket{\psi}=E^{B}_\mathrm{eff}\ket{\psi}$.
With some analytical calculations, it is easy to yield the single quasi-particle spectrum,
\begin{equation}
E^{\mathrm{B}}_\mathrm{eff}(K) =V+\frac{12J^2}{V}\cos^2\left(\frac{K}{2}\right).
\end{equation}
Similarly, from Eq.~(\ref{Eq_Eff_FH}), the two fermionic walkers and the two hard-core bosonic walkers obey the same effective single-particle Hamiltonian
\begin{equation}
  \hat H^{\mathrm{FH}}_\mathrm{eff} = J^\mathrm{FH}_\mathrm{eff} \sum_q\left(\hatd b_q\hat b_{q+1} +\hatd b_{q+1}\hat b_q\right) +\mu^\mathrm{FH}_\mathrm{eff} \sum_q\hatd b_q\hat b_q,
\end{equation}
but with $J^\mathrm{FH}_\mathrm{eff}=\slfrac{J^2}{V}$, $\mu^\mathrm{FH}_\mathrm{eff}=V+\slfrac{2J^2}{V}$ and spectra
\begin{equation}
E^{\mathrm{FH}}_\mathrm{eff}(K) =V+\frac{4J^2}{V}\cos^2\left(\frac{K}{2}\right).
\end{equation}
We observe that, for fixed values of $J$ and $V$, the hopping strength ($J^\mathrm{B}_\mathrm{eff},\, J^\mathrm{FH}_\mathrm{eff}$) of the composite particle essentially depend on their quantum statistics.
This means that quantum statistics has a significant effect on the co-walking of two interacting walkers.
In time-evolution dynamics, different values of hopping strength mean different walk speed.
Thus it is possible to explore statistic-dependent quantum co-walking via observing the walk dynamics.

\begin{figure}[!htp]
\includegraphics[width=1.0\columnwidth]{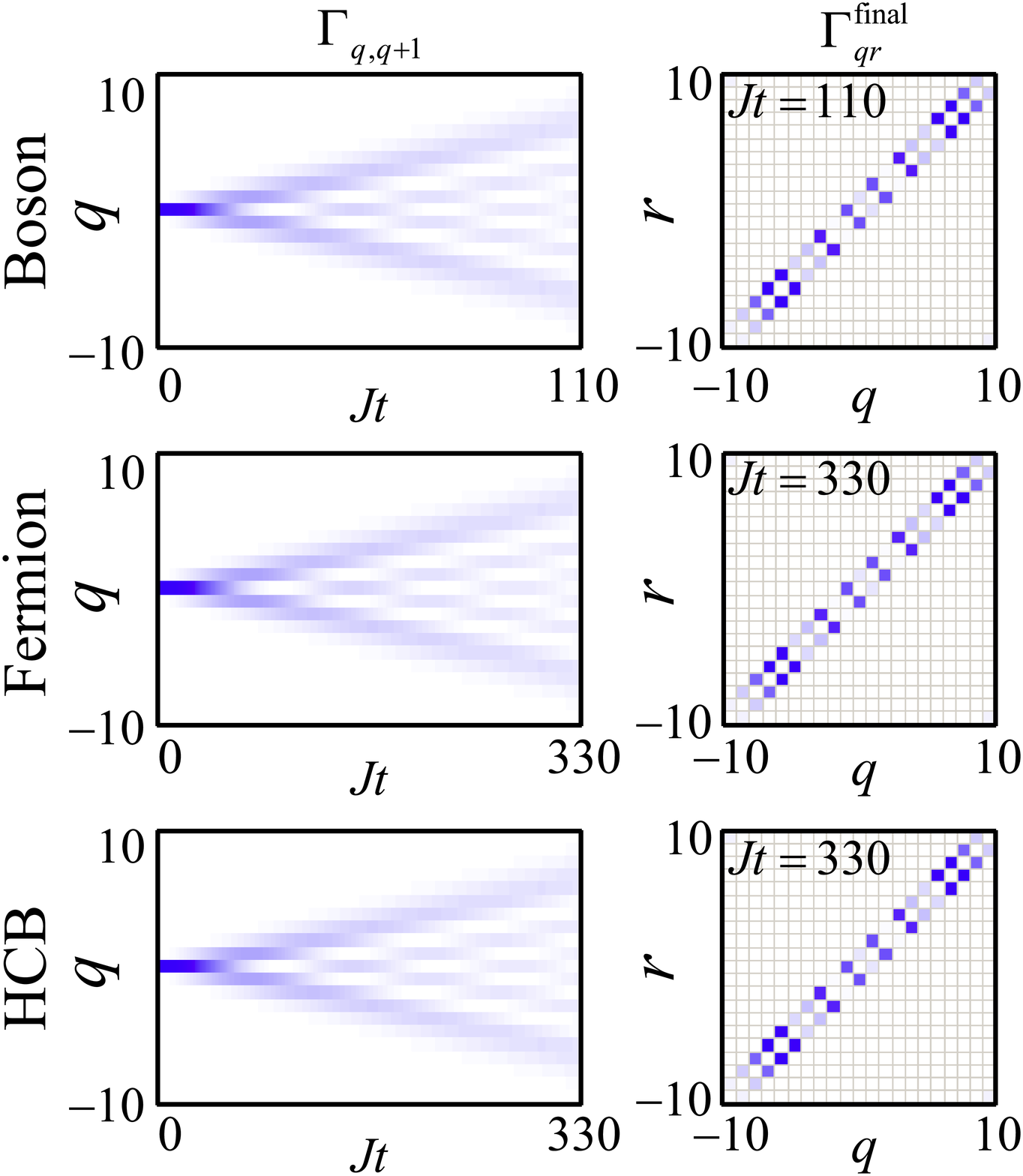}
\caption{\label{Fig_TPCF_Bound}(Color online) Quantum co-walking of two strongly interacting walkers with $\left|V/(2J)\right|=40$. Left: Time evolution of the minor diagonal correlations $\Gamma_{q,q+1}$. Right: Two-particle correlations $\Gamma_{q,r}^\mathrm{final}$ for the final states.}
\end{figure}

In Fig.~\ref{Fig_TPCF_Bound}, we show our numerical results for the time-evolution of the minor diagonal correlations $\Gamma_{q,q+1}$ and the final two-particle correlations $\Gamma_{q,r}^\mathrm{final}$ with $\left|V/(2J)\right|=40$.
From the correlations $\Gamma_{q,r}^\mathrm{final}$ in the right column of Fig.~\ref{Fig_TPCF_Bound}, we find that the two strongly interacting walkers are dominated by quantum co-walking.
From the time evolution of $\Gamma_{q,q+1}$ in the left column of Fig.~\ref{Fig_TPCF_Bound}, we see that the walk speed of two bosonic walkers is just three times of the ones of two fermionic and hard-core bosonic walkers.
These numerical results of spread are well consistent with our analytical prediction $J^\mathrm{B}_\mathrm{eff} = 3J^\mathrm{FH}_\mathrm{eff}$ from the second-order perturbation theory.

\section{Summary and discussion}\label{sec6}

In summary, we have explored how quantum statistics and inter-particle interactions affect two-particle QWs in 1D lattices with nearest-neighbor interactions.
Due to the inter-particle interactions, two particles with different quantum statistics undergo independent- and/or co-walking.
The QWs are dominated by independent-walking in the weak interaction limit, and vice versa, they are dominated by co-walking in the strong interaction limit.
We have analytically derived the effective single-particle model for the co-walking of two strongly interacting particles.
We find that the walk speed for the co-walking of two bosons is exactly three times of the ones for the co-walking of two fermions or two HCBs.
Although we only consider the two-particle QWs in attractive systems ($V<0$) in this article, similar QWs may be found in repulsive systems ($V>0$) which have free scattering states in the lower band and repulsively bound states in the upper band~\cite{Denschlag+Winkler}.
Our results for the case of two HCBs agree with the recent experimental observation of quantum dynamics of two atomic spin impurities~\cite{Fukuhara2013a}.
Besides observing bound states~\cite{Kitagawa2012,Fukuhara2013a}, our results of two-particle QWs provide promising applications in exploring quantum statistics.

\begin{figure}[!htp]
\includegraphics[width=1.0\columnwidth]{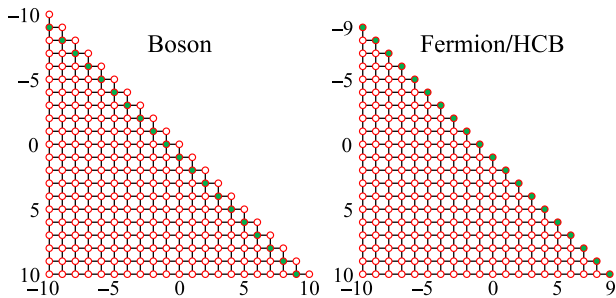}
\caption{\label{Fig_Waveguide}(Color online) Classical simulation with two-dimensional optical waveguide arrays. Each circle represents a waveguide. Green-colored and uncolored circles label waveguides with different refractive indices. The black lines connecting different circles denote their couplings. Left: the waveguide arrays for simulating two bosons. Right: the waveguide arrays for simulating two fermions or two hard-core bosons.}
\end{figure}

Furthermore, beyond a theoretical model, the two interacting quantum walkers in our models can be experimentally simulated with ultracold atoms in optical lattices and light waves in waveguides.
By using spin impurities of ultracold atoms in optical lattices,
two-magnon dynamics in the 1D Heisenberg XXZ chain has been observed in a recent experiment~\cite{Fukuhara2013a}.
It was a dramatic realization of two-HCB quantum walks with intermediate interaction ($\Delta=\left|V/(2J)\right|=0.986$).
The strong interaction regime ($\Delta \gg 1$) can be achieved by Feshbach resonance~\cite{Widera2004+Gross2010}.
Moreover, based on the quantum-optical analogues using engineered photonic waveguides~\cite{Longhi2009,Szameit2010}, the two-particle QWs obeying the Hamiltonian~(\ref{Eq_Hamiltonian}) can be simulated via light propagations.
As a single quantum walker in a 2D lattice is equivalent to two quantum walkers in a 1D lattice~\cite{Schreiber2012}, the two-particle QWs in 1D lattices can be simulated with light waves in 2D waveguide arrays~\cite{Szameit2009+Corrielli2013,Szameit2010}.
The temporal evolution of the superposition amplitude $C_{l_1l_2}$ in the two-particle Hilbert space is mapped onto the spatial propagation of the optical field $\mathbf{E}_{l_1l_2}$ in the $(l_1,l_2)$-th waveguide.
According to the evolution equation~(\ref{Eq_TE}) of $C_{l_1l_2}$, the propagation equation for $\mathbf{E}_{l_1l_2}$ is given by
\begin{eqnarray}
  i\frac{\mathrm{d}}{\mathrm{d} z} \mathbf{E}_{l_1l_2}&=&
    -J\left(\mathbf{E}_{l_1,l_2+1}+\mathbf{E}_{l_1,l_2-1}\right) \nonumber \\
  &&-J\left(\mathbf{E}_{l_1+1,l_2}+\mathbf{E}_{l_1-1,l_2}\right)\nonumber \\
  &&+V_{l_1 l_2}\mathbf{E}_{l_1l_2},
\end{eqnarray}
with $V_{l_1l_2}=V\delta_{l_1,l_2\pm 1}$ and the propagation distance $z$.
In Fig.~\ref{Fig_Waveguide}, we shown the 2D waveguide arrays for simulating two-particle QWs with $L_t=21$.
Similar to the 2D waveguide arrays used in recent experiments~\cite{Szameit2009+Corrielli2013,Szameit2010}, the waveguide arrays shown in Fig.~\ref{Fig_Waveguide} can be fabricated in a silica substrate by direct waveguide writing with femtosecond lasers~\cite{Gattass2008}.
Here the inter-particle interaction strength $V$ is controlled by the difference of refractive indices between green-colored and uncolored waveguides.

\acknowledgments
We thank Gora Shlyapnikov for discussion. This work is supported by the National Basic Research Program of China (NBRPC) under Grants No. 2012CB821305 and 2012CB922101, the National Natural Science Foundation of China (NNSFC) under Grants No. 11374375 and 11374331, and the Ph.D. Programs Foundation of Ministry of Education of China under Grant No. 20120171110022. XWG is partially supported by the Australian Research Council.

%


\end{document}